\documentstyle[eqsecnum,psfig,floats,twocolumn,aps,prb]{revtex}
\def\R{{\mathbf R}}
\def\rel{R}
\def\dyn{D}
\def\Re{{\bf{Re}}}
\def\Im{{\bf{Im}}}

\def\pvec{{\vec p}}
\def\ppvec{{\vec p'}}
\def\qvec{{\vec q}}

\def\jvec{\vec j}
\def\Avec{\vec A}

\begin{document}
\ifx\undefined\psfig\def\psfig#1{ }\else\fi
\ifpreprintsty\else
\twocolumn[\hsize\textwidth%
\columnwidth\hsize\csname@twocolumnfalse\endcsname
\fi
\draft
\preprint{ }
\title {The Elasticity of an Electron Liquid}
\author  {S. Conti}
\address {Max-Planck-Institute for Mathematics in the Sciences,
  04103 Leipzig, Germany}
\author  {G. Vignale}
\address{Department of Physics,  University of Missouri, Columbia,
  Missouri 65211}
\date{\today}
\maketitle
\begin{abstract}
The zero temperature response of an interacting
electron liquid to a time-dependent vector potential  of
wave vector $q$ and frequency $\omega$ such that $q\ll q_F$,
$qv_F\ll \omega \ll  E_F/\hbar$ (where $q_F$, $v_F$ and $E_F$ are the Fermi
wavevector, velocity
and energy respectively) is equivalent to that of a continuous
elastic medium with nonvanishing {\it shear modulus} $\mu$, bulk
modulus $K$, and  viscosity coefficients $\eta$ and $\zeta$.
We establish the relationship between the visco-elastic coefficients
and the long-wavelength limit of the ``dynamical
local field factors" $G_{L(T)}(q,\omega)$, which are widely used to describe
exchange-correlation effects in electron liquids.
 We present several exact results for $\mu$, including its expression in terms
of Landau parameters,  and practical approximate  formulas
for   $\mu$, $\eta$ and $\zeta$ as functions of density. These are used to
discuss the possibility of a {\it transverse} collective mode in the electron
liquid at sufficiently low density.
Finally, we consider impurity scattering and/or quasiparticle collisions at non
zero temperature.  Treating these effects in the
relaxation time ($\tau$) approximation, explicit expressions are derived for
$\mu$ and $\eta$ as functions of frequency.  These formulas exhibit a
crossover from the collisional regime ($\omega \tau \ll 1$), where $\mu
\sim 0$ and $\eta \sim nE_F\tau$, to the collisionless regime  ($\omega
\tau
\gg 1$), where $\mu \sim nE_F$ and $\eta \sim 0$.

\end{abstract}  
\ifpreprintsty\else\vskip1pc]\fi \narrowtext

\section{Introduction}
The  response of a {\it solid} body to  external macroscopic
forces is
described by the theory of elasticity.\cite{Landau7} In a homogeneous
and isotropic body \cite {Footnote0} the response is controlled by two
real elastic constants,  the bulk modulus $K$ and the shear modulus
$\mu$: dissipation is negligible.

The macroscopic response of a {\it liquid} system, on the other hand, is
usually described in terms of the Navier-Stokes equation \cite
{Landau6} of classical hydrodynamics. This is, at first sight, very
different from elasticity. First of all, by the very definition of liquid,
the shear modulus vanishes.  Second, there {\it is} dissipation, due to the
two viscosity coefficients $\eta$ and $\zeta$  -- the ``shear" and ``bulk"
viscosities  respectively. Only the bulk modulus remains approximately the same
in the liquid as in the solid state.

Such a sharp distinction  disappears at  finite
frequencies, where liquids develop solid-like characteristic, namely, a
non
vanishing shear modulus.
Both liquids and solids
 follow a common {\it visco-elastic} behavior,
which can be mathematically described by a single set of equations
(say the equations of elasticity)  with complex
frequency-dependent elastic constants
\begin {equation}
  \tilde K(\omega) = K(\omega) - i \omega \zeta (\omega)\,,
  \label {tildeK}
\end {equation}
and
\begin {equation}
  \tilde \mu(\omega) = \mu (\omega) - i \omega \eta(\omega)\,.
  \label {tildemu}
\end {equation}
The visco-elastic coefficients  $K$, $\mu$, $\zeta$, $\eta$  on the right
hand side are all real functions of frequency.

The crucial parameter that controls the prevalence of solid-like or
liquid-like behavior in the liquid is $\omega \tau$, the ratio of the
frequency to
the inverse of the relaxation time $\tau$ --
the  time it takes the system to return to
thermal equilibrium after being slightly disturbed from it. If $\omega
\tau \ll 1$ one is in the collision-dominated (or
hydrodynamic) regime, in which $\mu(\omega)$ is negligible and $\eta
(\omega)$, $\zeta (\omega)$ are finite.  If, on the other hand, $\omega
\tau \gg 1$ one is in the collisionless (or elastic) regime, where
$\mu(\omega)$ has a finite value, while the viscosities are small.
In either case, the bulk modulus does not show a significant
dependence on frequency.

In this paper we  explore the possibility of describing the long
wavelength dynamics of a quantum Fermi liquid \cite{PinesNoz} near the absolute
zero of temperature in terms of classical visco-elastic equations of motion.
Limiting ourselves to the {\it linear response} of the quantum liquid
to an external vector
potential $\vec A(q,\omega)$ of wave vector $q$ and frequency $\omega$ we
shall
show that  the visco-elastic description is possible (and useful) in
the regime
\begin {equation}  \label{equation1}
q \ll
q_F,~~~~~q \ll \omega/v_F, \end {equation}
where $v_F$ is the quasiparticle Fermi
velocity and $q_F$ is the Fermi wave vector. In other words, the frequency must
be high compared to the characteristic energy of quasiparticle-quasihole pairs
at wave vector $q$,  which tends to zero when $q \to 0$ (see Figure
\ref{figregime}).

\begin{figure}
  \psfig{figure=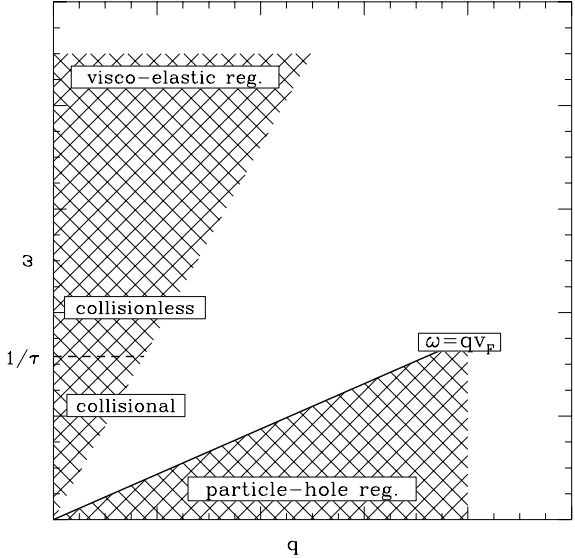,width=0.95\columnwidth}
  \caption{The region below the line $\omega=v_Fq$ is the
    quasiparticle-quasihole regime. The viscoelastic approach applies
    to the region $\omega \gg  v_Fq$. The line $\omega=1/\tau$ separates
the
    collisional viscoelastic regime and the collisionless viscoelastic
    one.}
  \label{figregime}
\end{figure}

The viscoelastic coefficients will be expressed in terms of the long wavelength
limit of the dynamical local field factors $G_{L(T)}(q,\omega)$: these are
mathematical constructs (defined below) that are widely used to describe
exchange-correlation effects in Fermi liquids.

Most of this paper is devoted to the task of {\it calculating} the
viscoelastic coefficients of an electron liquid (both in three and two
dimensions)  in the limit of $\omega \to 0$, that is, in practice, for  $\omega
\ll E_F$ but still satisfying the condition~(\ref{equation1}).
Such coefficients are particularly relevant in the framework of
Time-Dependent Density Functional Theory,\cite{td-dft} where they
fully determine the low-frequency regime.

We first consider the case of a  uniform (translationally invariant) electron
gas at the absolute zero of temperature. In this case the low-frequency
elastic
constants $K$ and $\mu$ can be expressed exactly in terms of the Landau
parameters $F_0$, $F_1$ and $F_2$ -- at least insofar as the Landau theory of
Fermi liquids applies.  The result for the bulk modulus [$K=n^2d^2 \epsilon
(n)/dn^2$, where $\epsilon(n)$ is the ground-state energy density and $n$
is the
particle density] has been known for a long time  \cite{PinesNoz,Nozieres} and
can be straightforwardly evaluated  from the knowledge of the ground
state energy.
\cite{CeperleyAlder,Tanatar89}  The result for the shear modulus is (to
the best of our knowledge) new and,  unfortunately, not so easy to evaluate.
For this reason, we propose an approach ``\`a la Wigner", namely we
calculate the
shear modulus at both high and low densities -- where the calculation
can be done
with relative ease -- and interpolate  between these two limits.  The proposed
interpolation function  is close to the results of recent mode-coupling
calculations of the dynamical local field factor\cite {Nifosi} at sufficiently
high density.

We proceed in a similar way to the calculation of the viscosities.  First the
shear viscosity $\eta$ is analytically calculated in the high density limit,
making use the formalism of Nifos\'\i\ {\em et al.},\cite {Nifosi} for the
imaginary part of the dynamical local field factor.  Then we devise a numerical
fit that reduces to the analytical result in the high density limit and
reproduces the numerical data of Nifos\'\i\ {\em et al.},\cite
{Nifosi} at lower
density.
 The bulk viscosity $\zeta$ is found to be approximately zero in this
approach.

The last part of the paper is devoted to the treatment  of relaxation effects
caused either by collisions with impurities or by collisions between
thermally excited quasiparticles.
 We
assume that both effects can be described by a single relaxation time $\tau$,
such that $1/\tau \ll E_F$, and  make use of the standard ``relaxation time
approximation" (RTA) \cite {Mermin70Das75} to approximate the
collision integral in the quasiparticle transport equation.  The low-frequency
regime now splits into two distinct regimes: collisional ($\omega \tau \ll 1$)
and collisionless ($\omega \tau \gg 1$).  The restriction given by
Eq.~(\ref{equation1}) remains in force in both regimes.
By solving the transport equation in the RTA we obtain explicit expressions
for the elastic and viscous coefficients.  In the collisionless regime, these
expressions reduce to the ones derived in the first part of the paper.  In
the collisional regime, they are very different:  the shear modulus
vanishes (as expected for an ordinary liquid), and the shear viscosity
tends to the limit $\eta = \mu \tau$  where $\mu$ is the collisionless
shear
modulus. Our simple analytic expressions clearly exhibit the crossover from
the collisional to the collisionless regime.\cite{heliumnote}

We have emphasized the importance of the condition ~(\ref{equation1}) that
assures the possibility of a visco-elastic description of the dynamics of the
Fermi liquid.  What happens if this condition is violated?  The behavior of
the microscopic current-current response function of a Fermi liquid changes
dramatically as one goes from the $q \ll \omega/v_F$ regime to the
$q \gg \omega/v_F$ regime, even though $q$ and $\omega$ remain small
compared to $q_F$ and $E_F$ respectively.  The physical reason is that the
response in the second  region is dominated by  electron-hole
excitations that are absent in the first.  Because of the change in the
character of the response, the current {\it does not} obey classical
visco-elastic equations of motion in the second regime. Alternatively,  if
one insisted on casting the equation for the current in a visco-elastic
form, one would be forced to use visco-elastic coefficients that {\it diverge}
in the $q \to 0$  limit.   This shows that the visco-elastic theory is not a
natural description of the physics for $\omega < qv_F$.

This paper is organized as follows:

In Section \ref{secII} we briefly review elasticity, hydrodynamics, and
the local field factor representation of the current-current  response
functions
of a Fermi liquid.  We  establish the relationship  between the dynamical
local field  factors and the frequency-dependent  visco-elastic coefficients of
Eqs.~(\ref{tildeK}-\ref{tildemu}).

In Section \ref{secIII} we derive an exact expression for the
shear modulus of the Fermi liquid at $T=0$ in term of Landau parameters, and a
rigorous upper bound on the value of the elastic constants.

In Section \ref{secIV} we present approximate analytical expressions for the
evaluation of the visco-elastic coefficients of  an interacting electron
liquid as functions of density. These expressions  are used to discuss the
possibility of a {\it transverse} sound mode in the low density electron gas.

In Section \ref{secV} we include electron-impurity  and thermally induced
quasiparticle collisions  {\it via} Mermin's relaxation time approximation. We
provide explicit formulas for the frequency dependent (on the scale of the
inverse relaxation time) shear modulus and  viscosity, exhibiting the crossover
between the collisional and the collisionless regimes.

\section{Viscoelastic constants of a Fermi liquid}
\label{secII}
\subsection {Macroscopic equations}
The
equation of motion for the elastic displacement field $\vec u(\vec r,t)$
in a homogeneous and isotropic solid is \cite{Landau7}
\begin {equation} mn {\partial^2 \vec u  \over \partial t^2} =
  \left[K + \left(1-{2 \over d}\right)\mu \right]
  \vec \nabla(\vec \nabla \cdot \vec u) +
  \mu \nabla^2 \vec u + \vec F(\vec r,t) \label {elasticmotion}
\end{equation}
where $K$ and $\mu$ are constants, known as the bulk and
the shear modulus respectively, $d$ is the number of space dimensions,
$\vec F(\vec r,t)$ is an externally applied volume force density, $n$
is the equilibrium number density, and $m$ is the mass of the particles.

We consider periodic forces of the form
\begin{equation}
  \vec F(\vec r,t) = \vec F (\vec q,\omega) e^{i(\vec q \cdot \vec
    r - \omega t)} + c.c., \label{F1}
\end {equation}
which induce periodic displacements
\begin {equation}
  \vec u (\vec r,t) = \vec u(\vec q,\omega)e^{i(\vec q \cdot \vec r -
    \omega t)} +
  c.c. \label{uqw}
\end {equation}

In order to make contact, later, with microscopic theories of Fermi
liquids, we write the force as the time derivative of a vector potential
\cite{Footnote00}
\begin {equation}
  \vec F(\vec r,t) = n {\partial \vec A(\vec r,t) \over \partial t}
  \label {A1}
\end {equation}
and introduce the current density
\begin {equation}
  \vec j(\vec r,t) = n {\partial \vec u(\vec r,t) \over \partial t}
\label {j1}
\end {equation}
as its conjugate field. The equation of motion (\ref{elasticmotion}),
written in terms of the Fourier transform of the current density, takes
the form
  \begin{eqnarray}
-i \omega m \vec j(\vec q, \omega) &=& \left[{K \over n} + \left(1 - {2
      \over
d}\right) {\mu \over n} \right] {\vec q [\vec q \cdot \vec j (\vec
      q,\omega)] \over i
\omega}\nonumber\\
&& + {\mu \over n}{q^2 \over i \omega} \vec j(\vec q,\omega) - i
\omega n \vec A(\vec q,\omega). \label{elasticmotion2}
  \end{eqnarray}
Both the current and the vector potential can be written as sums of
longitudinal and transverse components (parallel and perpendicular to
$\vec q$ respectively) $\vec j = \vec j_{L} + \vec j_{T}$,
$\vec A = \vec A_{L} + \vec A_{T}$ and the equations of motion for
longitudinal and transverse components decouple.  The solution of
Eq.~(\ref{elasticmotion2}) is
\begin {equation}
  \vec j_{L}(\vec q,\omega) = {n/m \over 1 - \left[ {K \over n^2} +
      2\left(1 - {1\over
          d}\right){\mu \over n^2} \right] {nq^2 \over m \omega^2}} \vec
  A_{L} (\vec
  q, \omega)
  \label{elasticresponseL}
\end {equation}
and
\begin {equation}
  \vec j_{T}(\vec q,\omega) = {n/m \over 1 - {\mu \over n^2}{nq^2 \over m
      \omega^2}} \vec A_{T} (\vec q,
  \omega). \label{elasticresponseT}
\end{equation}
We note that $\vec j_{L}(\vec q,\omega)$ is related to the  induced density
change $n_1(\vec q,\omega)$ by the continuity equation
\begin {equation}
  n_1(\vec q,\omega) = {\vec q \over \omega} \cdot \vec j_{L} (\vec q,\omega)
  \label{continuity}
\end {equation}
and the longitudinal vector potential
$\vec A_{L}(\vec q,\omega)$ is equivalent, {\it modulo} a gauge
transformation,
to a {\it scalar} potential $V(\vec q, \omega)$ such that
\begin {equation}
  \vec A_{L}(\vec q,\omega) = {\vec q \over \omega} V(\vec q,\omega).
  \label{gaugeinvariance}
\end {equation}

In writing these equations we have assumed that the external field $\vec
A_{L}$ includes, self-consistently, the contribution of the
mean electrostatic field generated by the density fluctuation
$n_1$  [the Hartree field $V_H =
v(q) n_1$].

Let us now consider the classical hydrodynamical equation for the current
density in a liquid.\cite {Landau6} In the linear approximation with respect to
$\vec j$ and $\vec A$ it has the form  \begin {eqnarray}
  -i \omega m \vec j(\vec q, \omega) &=& \left[ {K \over n}-{i \omega
      \zeta \over n} + \left(1 - {2 \over d}\right) {i \omega \eta
      \over n} \right] {\vec q [\vec q \cdot
    \vec j (\vec q,\omega)] \over i \omega} \nonumber\\
  &&+ {i \omega \eta \over n}{q^2 \over i
    \omega} \vec j_1(\vec q,\omega) - i \omega n \vec A(\vec q,\omega),
  \label{hydromotion2}
\end{eqnarray}
where we have used the continuity equation ~(\ref{continuity}) to
rewrite the hydrostatic pressure term in the Euler equation\cite
{Landau6} as
\begin {equation}
  - \vec \nabla p (\vec q,\omega) = - {dp(n) \over dn} \vec \nabla
  n_1(\vec q, \omega) = {dp(n) \over dn}  \vec q \left[{\vec q \cdot
      \vec j_(\vec q,\omega) \over i \omega}\right].
\label{hydroforce}
\end {equation}
The constants $\eta$ and $\zeta$ are  the shear and bulk
viscosity coefficients respectively; $p(n)$ is the equilibrium pressure
as a function of density and is related to the bulk modulus by the
relation $K = n dp(n)/dn$.

Equations ~(\ref{hydromotion2}) and ~(\ref
{elasticmotion2}) are  very similar. The hydrodynamic
equation differs from the elastic equation through the following
replacements:  (i) The shear modulus $\mu$ is replaced by the imaginary
quantity $-i \omega \eta$, which vanishes at $\omega = 0$ in agreement
with the notion that a liquid has no resistance to shear.   (ii) The bulk
modulus $K$ acquires an imaginary part $- i \omega \zeta$ where $\zeta$ is the
bulk viscosity.

These observations suggest the use of a single language - say that of
elasticity theory - to describe both the liquid and the solid.  In this
generalized scheme, the equation of motion becomes
\begin {eqnarray}
  -i \omega m \vec j(\vec q, \omega) &=& \left[{\tilde K(\omega) \over n}
    + \left(1 - {2 \over d}\right) {\tilde \mu (\omega) \over n}\right]
  {\vec q [\vec q \cdot \vec j (\vec q,\omega)] \over i
    \omega}\nonumber\\
  &&+
  {\tilde \mu (\omega) \over n}{q^2 \over i
    \omega} \vec j(\vec q,\omega) - i \omega n \vec A(\vec q,\omega),
  \label{viscoelasticmotion}
\end{eqnarray}
where $\tilde \mu (\omega)$ and $\tilde K(\omega)$ are the
frequency-dependent visco-elastic constants defined in the introduction
[Eqs.~(\ref{tildeK}-\ref{tildemu})].

The solution of this equation of motion is
\begin {equation}
  \vec j_{L}(\vec q,\omega) = {n/m \over 1 - \left[{\tilde K (\omega)
        \over n^2} +
      2\left(1 - {1\over d}\right){\tilde \mu (\omega) \over n^2}\right]
    {nq^2 \over m \omega^2}}
  \vec A_{L} (\vec q, \omega) \label{viscoelasticresponsel}
\end {equation}
and
\begin {equation}
  \vec j_{T}(\vec q,\omega) = {n/m \over 1 - {\tilde \mu (\omega) \over
      n^2}{nq^2 \over m \omega^2}} \vec A_{T} (\vec q, \omega).
  \label{viscoelasticresponset}
\end {equation}

The key difference between a solid and a liquid is that the solid has an
essentially real $\tilde \mu$ ($\mu$ finite, $\eta \sim 0$), whereas a liquid
has an essentially imaginary $\tilde \mu$ ($\eta$ finite, $\mu \sim 0$). As for
the generalized bulk modulus, its real part (related to the compressibility) is
nearly the same in the two phases. The bulk viscosity is generally of
the same order of magnitude as the shear viscosity,\cite{Landau6}  but, in the
case of the electron liquid, it will be shown to  vanish within the
mode-coupling
approximation of Ref.~\onlinecite{Nifosi}.

\subsection{ Connection with microscopic linear response theory}
Let us now turn to the microscopic formulation of the linear response of
a homogeneous, isotropic body, subjected to an external vector
potential $\vec A(\vec q, \omega)$.  The {\it proper} response
functions $\chi_L(\vec q, \omega)$ and $\chi_T(\vec q, \omega)$
(longitudinal and transverse respectively) are defined by the relations
\begin {equation}
\vec j_{L(T)} (\vec q, \omega) = \chi_{L(T)} (\vec q,\omega)
\vec A_{L(T)} (\vec q, \omega). \label {linearresponse} \end{equation}

A useful way of representing  $\chi_{L(T)}$ is \cite{SingwiTosi}
\begin {equation}
\chi_{L(T)}(\vec q, \omega) =  {\chi^0_{L(T)}(\vec q,\omega) \over 1+
v(q)G_{L(T)}(\vec q,\omega) (q^2/\omega^2)\chi^0_{L(T)}(\vec q,\omega)},
\label{chi} \end{equation}
where
\begin {equation}
\chi^0_{L(T)}(\vec q,\omega) = {n \over m} + \sum_{\vec k} \left(
{k_{L(T)} \over m}\right)^2 {f(\epsilon_{k+q}) - f(\epsilon_k) \over
\omega + \epsilon_{k+q}-\epsilon_k} \label{chi0s} \end {equation} are the
longitudinal (transverse) response functions of the  noninteracting
electron gas, $\epsilon_k = k^2/2m$
is the free particle energy, $f(\epsilon_k)$ is the Fermi distribution
function, and $k_{L(T)}$ is the longitudinal (transverse) component of
$\vec k$ relative to $\vec q$, and $v(q)$ is the Fourier transform of the
interaction [$v(q)=4 \pi e^2/q^2$ and $v(q)=2 \pi e^2/q$ in three and two
dimensions respectively]. The dynamical local field factors  $G_{L(T)}(\vec
q,\omega)$ are effectively {\it defined}  by these equations: they take into
account exchange-correlation effects beyond the random phase approximation.

Note that, with these definitions,  the longitudinal local field factor $G_L$
coincides with the more familiar scalar local field factor $G$
used in the theory of the density-density response function in
Ref.~\onlinecite {SingwiTosi} for example.


Let us now consider the  long wavelength limit
($qv_F\ll \omega$) of $\chi_{L(T)} (q,\omega)$. Expanding the
noninteracting
response functions ~(\ref{chi0s}) we find
 \begin {eqnarray}
\chi_{L(T)}(\vec q, \omega) &\simeq& {n \over m}\left[1+
  \alpha_{L(T)}^{(d)}{E_F
\over n} {nq^2 \over m \omega^2} \right.\nonumber\\
&& \hskip1cm\left.+ f_{xcL(T)}(\omega){nq^2 \over m
\omega^2}\right], \label{chismallq} \end {eqnarray}
where
\begin {equation}
  f_{xcL(T)}(\omega) \equiv - \lim_{q \to 0} v(q) G_{L(T)}(q,\omega)
  \label{deffxc}
\end {equation}
are complex functions of frequency, which satisfy Kramers-Kr\"onig
dispersion relations between their real and imaginary parts  (see discussion
in the next section) and $\alpha_L^{(3)} =6/5$, $\alpha_L^{(2)} =3/2$,
$\alpha_T^{(3)} =2/5$, $\alpha_T^{(2)} =1/2$.

Comparing Eq.~(\ref{chismallq}) with the
macroscopic response functions ~(\ref{viscoelasticresponsel}) and
 ~(\ref{viscoelasticresponset}) we are led to the following
identifications:
\begin {equation}
\tilde \mu (\omega) = \alpha_T^{(d)} nE_F +n^2 f_{xcT}(\omega)
\label{mutildemicro} \end {equation}
and
\begin {eqnarray}
  \tilde K (\omega) &=& \left[\alpha_L^{(d)}-\left(2 -{2\over d}
    \right)\alpha_T^{(d)} \right]nE_F \nonumber\\
  &&+n^2 \left[ f_{xcL}(\omega) - \left(2-{2 \over d}
    \right)f_{xcT}(\omega) \right]\,.
  \label{Ktildemicro}
\end {eqnarray}

Separating the real and imaginary parts of these equations, and taking
the limit $\omega \to 0$ (but still with $\omega \gg qv_F$), we arrive at the
promised expressions for the elastic and viscosity coefficients in terms of the
long wavelength limit of the local field factors: \begin {equation}
  \mu = \alpha_T^{(d)}nE_F +n^2 {\bf Re} f_{xcT}(0),
  \label{eq1}\end {equation}
\begin {eqnarray}
  K&=& \left[\alpha_L^{(d)}-\left(2-{2\over d}\right) \alpha_T^{(d)}
  \right]nE_F
  \nonumber\\
  &&+n^2 {\bf Re} \left[f_{xcL}(0)-\left(2-{2\over d}\right) f_{xcT}(0)
\right],
  \label{eq2}
\end {eqnarray}
\begin {equation}
  \eta = -n^2 \lim_{\omega\to 0} {{\bf Im} f_{xcT}(\omega)\over \omega},
  \label{eq3}
\end {equation}
\begin {equation}
  \zeta = -n^2\lim_{\omega\to 0}\left({{\bf Im} f_{xcL}(\omega)\over\omega}-
    \left(2-{2\over d}\right){\Im
      f_{xcT}(\omega) \over \omega}\right] \label{eq4}
\end {equation}

Equations ~(\ref{eq1}--\ref{eq4}) are the
main result of this Section. We underline the fact that they have been
obtained under the assumption  $q v_F \ll \omega$. Outside this regime, e.g.,
for $\omega < qv_F$,  the microscopic response functions do not yield
visco-elastic equations of motion or, equivalently,  the visco-elastic
coefficients diverge for $q \to 0$.  The analysis of this non-viscoelastic
regime is beyond the scope of this paper.

\section{The calculation of the elastic constants: rigorous results}
\label{secIII}
\subsection {Expression in terms of Landau parameters}
The elastic constants of a Fermi liquid can be exactly expressed in terms of
Landau parameters, insofar as the Landau theory of Fermi liquids is valid.
To
see this, we begin by deriving the equation of motion for the quasiparticle
distribution function  in the presence of a slowly varying {\it vector}
potential
$\vec A (\vec r,t)$.

Following the discussion of Nozi\`eres and Pines \cite
{PinesNoz} we treat the quasiparticles as a gas of non-interacting classical
particles governed by a self-consistent Hamiltonian
\begin{eqnarray} \label{hqp}
H_{qp}(\vec r, \vec p) &=& \epsilon_{\vec p + \vec A (\vec r,t)} +
\sum_{\ppvec}
f_{pp'} \delta n_{p'} (\vec r, t) \nonumber \\ & \simeq& \epsilon(\vec
p)+{\vec p
\cdot \vec A(\vec r,t) \over m^*} +\sum_\ppvec
f_{pp'} \delta n_{p'}(\vec r, t), \end {eqnarray}
where $\epsilon_{\vec p}$ is the quasiparticle energy, $\vec \nabla_p
\epsilon_{\vec p} \equiv \vec v_p = \vec p /m^*$ is the quasiparticle
velocity,
$m^*$ is the
 effective mass, and $f_{pp'}$ are Landau parameters. This
approach is justified in the limit of zero temperature and zero frequency,
since
the excited quasiparticles have an essentially infinite lifetime in this
regime,
and their mutual collisions are negligible.

The
self-consistent nature of the quasiparticle Hamiltonian is apparent in the last
term, which is proportional to  the departure of the quasiparticle phase space
distribution function $n_p(\vec r,t)$ from the {\it local} equilibrium
distribution $n_{0} (\epsilon_{\vec p + \vec A})$:
\begin {eqnarray} \label {deltanqp}
\delta n_p (\vec r, t) &=& n_p(\vec r, t) -n_{0} (\epsilon_{\vec p + \vec
A}) \nonumber\\ & \simeq& n_p(\vec r, t) - n_{0}(\epsilon_p) -
n_0'(\epsilon_p) {\vec p \cdot \vec A(\vec r,t) \over m^*} \nonumber \\
&=& n_{1p} - n_0'(\epsilon_p)
{\vec p \cdot \vec A(\vec r,t) \over m^*}. \end {eqnarray}
Here $n_0 (\epsilon_p) = \theta (\epsilon_{p_F} - \epsilon_p)$ is the true
equilibrium distribution function at $T=0$ and chemical potential
$\epsilon_{p_F}$,
$n_0'(\epsilon_p) = - \delta(\epsilon_{p_F} - \epsilon_p)$ is its
derivative with
respect to energy, and
\begin {equation} \label {truedeparture}
n_{1p}(\vec r,t) = n_p(\vec r, t) - n_0 (\epsilon_p)
\end {equation}
is the departure of the distribution function from {\it true} equilibrium.

The fact that the departure from local rather than true equilibrium
appears in Eq.~(\ref{hqp}) is essential to guarantee particle conservation and
gauge invariance of the theory.

We  now make use of the well-known Landau relation \cite
{PinesNoz}  between bare and
effective mass in a translationally invariant system
\begin {equation} \label {effmass}
{\vec p \over m} = {\vec p \over m^*} - \sum_\ppvec
f_{pp'}n_0'(\epsilon_p') {\vec p' \over m^*}, \end {equation}
and  rewrite the effective Hamiltonian in terms of the departure from true
equilibrium [see Eq.~(\ref{deltanqp})]:
\begin {equation} \label{hqp2}
H_{qp}(\vec r, \vec p) = \epsilon(\vec p)+{\vec p
\cdot \vec A(\vec r,t) \over m} +\sum_\ppvec
f_{pp'} n_{1p'}(\vec r, t). \end {equation}

Finally, we write the classical (linearized) Liouville equation for the
evolution of the quasiparticle distribution function under  $H_{qp}$.  After
introducing the Fourier representation

\begin {equation} n_{1p}(\vec r, t) =
n_{1p}(\vec q,\omega) e^{i(\vec q \cdot \vec r - \omega t)} +c.c \label
{deltan}
\end {equation}
we obtain the desired  equation of motion
 \begin {eqnarray} \label {transportequation} &&(\vec q \cdot
\vec v_p - \omega) n_{1p}(\vec q,\omega)
\\&&-\vec q \cdot \vec v_p n_0'(\epsilon_p) \left[\sum_{p'}f_{pp'}n_{1p'}(\vec
q,\omega) + {\vec p \over m}\cdot \vec A (\vec q,\omega)\right] =0.
\nonumber\end {eqnarray}

The current response is
obtained from the quasiparticle distribution function via the relation
\begin{equation}
\vec j(\vec q, \omega) = \sum_p { \vec p \over m} n_{1p} (\vec
q,\omega)+ {n \over m} \vec A_1(\vec q,\omega),
\label{Landaucurrent} \end {equation}
where it must be noted that the bare mass, rather than the effective
mass,  enters the definition of the current.\cite{PinesNoz} The density
response  is given  by $n_1(\qvec,\omega) = \sum_p
n_{1p} (\qvec,\omega)$.

Equation (\ref{transportequation}) can be solved for a given value of the ratio
$x \equiv q v_F/\omega$, with both $q$ and $\omega$ tending to zero.
After setting \begin {equation} \label{Pip} n_{1p}(\qvec,\omega) = \vec
\Pi_\pvec
(x) \cdot \vec A(\qvec,\omega), \end {equation}
we see that $\vec \Pi_\pvec (x)$  obeys
the equation of motion
\begin{equation}\label{eqmotolandaupirp}
\vec \Pi_\pvec (x) = R_\pvec (x) \left( {\pvec \over m}  + \sum_\ppvec
f_{\pvec\ppvec} \vec \Pi_\ppvec (x) \right) \end{equation}
where
\begin{equation}\label{eqdefrp}
R_\pvec (x)  \equiv
n_0'(\epsilon_p){x \cos (\theta) \over
x \cos (\theta) -  1},
\end{equation}
and $\theta$ is the angle between $\pvec$ and $\qvec$.

In the $x\to 0$ limit [see Eq.~(\ref{equation1})] we expand $R_\pvec (x)$
and $\vec \Pi_\pvec (x)$ in  a power series of $x$ as follows

\begin {equation} \label {rexpansion}
R_\pvec (x) = -n_0'(\epsilon_p) \sum_{n=1}^\infty (x \cos {\theta})^n
\end{equation}
and
\begin {equation} \label {piexpansion}
\vec \Pi_\pvec (x) = \sum_{n=0}^\infty \vec \Pi_\pvec^{(n)} x^n.
\end{equation}

Inserting these expansions in Eq.~(\ref{eqmotolandaupirp}) we obtain the
recursion relation
\begin {eqnarray} \label {recursion}
\vec \Pi_\pvec^{(n)} &=& -n_0'(\epsilon_p) {\vec p \over m}
(\cos {\theta})^n \nonumber \\&& - n_0'(\epsilon_p) \sum_\ppvec f_{\pvec
\ppvec}
\sum_{m=0}^{n-1} \vec \Pi_\ppvec^{(m)} (\cos {\theta})^{n-m},
\end {eqnarray}
with $\vec \Pi_\pvec^{(0)} = 0$ and $\vec \Pi_\pvec^{(1)} =
-n_0'(\epsilon_p) \vec p \cos{\theta}/m$.
The relevant term is the one with $n=2$:
\begin {eqnarray} \label{Pi2}
\vec \Pi_\pvec^{(2)} &=& - n_0'(\epsilon_p) \left[ {\vec p
\over m} (\cos{\theta})^2 \nonumber \right. \\&&
\left. -\sum_\ppvec f_{\pvec \ppvec}\cos{\theta} n_0'(\epsilon_{p'}) {\ppvec
\over
m} \cos{\theta '} \right].  \end {eqnarray}
This enables us to calculate the current, and hence the response function,
exactly to order $x^2 = (q v_F/\omega)^2$.
More precisely,  making use of Eq.~(\ref{Landaucurrent}) we obtain
(from now on, we focus on 3D)
\begin {eqnarray} \label{chilnew} \chi_L  - {n \over m} &=& \sum_\pvec
{\hat q
\cdot \pvec
\over m}\hat q \cdot \vec \Pi_\pvec^{(2)} (x) x^2 \nonumber \\ &=&
{nq^2 \over m \omega^2}{q_F^2 \over m^2} {3/5+ 4F_2/75 + F_0/3 \over
1+F_1/3} \end {eqnarray}
and
\begin {eqnarray} \label {chitnew} \chi_L+2\chi_T -3{n \over
m} &=& \sum_\pvec
{ \pvec
\over m}\cdot \vec \Pi_\pvec (x) x^2 \nonumber \\ &=&
{nq^2 \over m \omega^2}{q_F^2 \over m^2}
{1+2 F_2 /15 + F_0/3 \over 1+F_1/3},
\end {eqnarray}
where $F_l$ are the usual dimensionless Landau parameters.\cite{PinesNoz}
The following partial results have been used to evaluate the sums over
$\pvec$  and $\ppvec$ in three dimensions

\begin {eqnarray} && \sum_{\pvec \ppvec} n_0'(\epsilon_p) n_0'(\epsilon_{p'})
f_{\pvec \ppvec} (\hat q \cdot \pvec)^2(\hat q \cdot \ppvec)^2 =
\nonumber \\&& {q_F^4 N(0) \over 9} \left(F_0+{4 \over 25}F_2 \right) \end
{eqnarray} and
\begin {eqnarray} &&\sum_{\pvec \ppvec} n_0'(\epsilon_p) n_0'(\epsilon_{p'})
f_{\pvec \ppvec} (\hat q \cdot \pvec)(\hat q \cdot \ppvec)(\pvec
\cdot \ppvec)= \nonumber \\&& {q_F^4 N(0)
\over 9} \left(F_0+{2 \over 5}F_2 \right) \end {eqnarray}
where $N(0) = m^*q_F/\pi^2$ is the three-dimensional density of
quasiparticle states at the Fermi surface.

A direct comparison between Eqs.~(\ref{chilnew}, \ref{chitnew}) and
Eq.~(\ref{chismallq}) yields the desired expressions for
$f_{xcL(T)}(0)$:

\begin {equation}
\Re f_{xcL} (0) = {6 E_F \over 5 n} {{5 \over 9} F_0+{4 \over
45}F_2-{1\over 3}F_1 \over 1+{1 \over 3}F_1} \label{RefxcL0} \end
{equation} and
\begin {equation}
\Re f_{xcT} (0) = {2 E_F \over 5n}{{1\over5}F_2-{1\over3}F_1 \over 1
+{1\over3}F_1}
\label{RefxcT0} \end {equation}

Similar results are obtained in {\it two} dimensions:
\begin {equation}
\Re f_{xcL} (0) = {2 E_F \over  n} {{1 \over 2} F_0-{3 \over
8}F_1+{1\over 8}F_2 \over 1+{1 \over 2}F_1}\,, \label{RefxcL02d} \end
{equation}
\begin {equation}
\Re f_{xcT} (0) = {2 E_F \over n}{{1\over8}F_2-{1\over8}F_1 \over 1
+{1\over2}F_1}
\label{RefxcT02d} \end {equation}

Substituting in Eqs.~(\ref{eq1}) and (\ref{eq2}) we obtain the following
expressions for the elastic constants.  In three dimensions
\begin {equation}
K = n^2 {d^2 \epsilon(n) \over dn^2} = {2 nE_F \over 3}{1+F_0 \over
1+{1\over3}F_1}
\label{K} \end {equation}
and
\begin {equation}
  \mu  = {2 nE_F \over 5}{1+ F_2/5 \over 1 +F_1/3} .
  \label {mu}
\end {equation}
In two dimensions
\begin {equation}
K  ={nE_F}
{1+F_0  \over 1+{1\over2}F_1}\,.
\label {compressibility2d} \end {equation}
and
\begin {equation}
  \mu  = { nE_F \over 2}{1+ F_2/2 \over 1 +F_1/2} .
  \label {mu2d}
\end {equation}

These are the main results of this Section.  There is no surprise as far
as the bulk modulus is concerned: it is given by the standard
thermodynamic expression, where the energy density can be calculated by
Quantum Monte Carlo.  The shear modulus, on the other hand, has an
expression involving Landau parameters, which are not easily calculated
from Monte Carlo simulations, even though some progress in this direction
has recently been reported.\cite{Kwon94}

\subsection {Rigorous upper bounds on the elastic constants}

In this Section we derive two exact bounds on the
elastic constants, which follow from the Kramers-Kronig dispersion
relations between the real and the imaginary parts of $f_{xcL(T)}
(\omega)$.
The origin of these relations can be easily seen from the formula
\begin {equation}  \label{deffxcnew}
f_{xcL(T)}(\omega) = \lim_{q \to 0} {\omega^2 \over
q^2} \left( [\chi^{(0)}_{L(T)}]^{-1}(q,\omega) -
\chi_{L(T)}^{-1}(q,\omega)  \right) \end {equation}
which directly follows from the representation ~(\ref{chi}) and the
definition ~(\ref{deffxc}). Both $\chi$ and $\chi^{(0)}$ are analytic
functions of frequency in the upper half plane of this variable, and both
have no zeroes in this domain.\cite{Landau5}  This implies that their
inverses are also analytic everywhere in the upper half
plane.  The large frequency behavior of Eq.~(\ref{deffxcnew}) is  regular
because $\chi$ and
$\chi^{(0)}$  have the same form  [$n/m + O(q^2/\omega^2$)] in this limit.
{}From this, one can also see that the $q \to 0$ limit is well behaved.

{}From these considerations we  conclude that the
Kramers-Kr\"onig relations must hold in the standard form
\begin {eqnarray}
\Re f_{xcL(T)}(\omega) &=& \Re f_{xcL(T)}(\infty)
\nonumber\\
&&+ {2 \over \pi} {\cal P}
\int_0^\infty d \omega' { \omega' \Im f_{xcL(T)}(\omega') \over
(\omega')^2 - \omega^2} \label {KramersKronig} \end{eqnarray}
where ${\cal P}$
denotes the Cauchy principal part.  The second law of thermodynamics
(positivity of dissipation) requires that ${\bf Im} \chi_{L(T)}
(q,\omega) \leq 0$ at all positive frequencies.  But
${\bf Im}
\chi^{(0)}_{L(T)}
(q,\omega) = 0$ for $q v_F \ll \omega$ [see Eq.~(\ref{chi0s})].
Thus, from Eq.~(\ref{deffxcnew}), we see  that
\begin {equation}
\Im f_{xcL(T)}(\omega) \le 0 \label{secondlaw}
\end{equation}
for all
positive frequencies. It then follows from Eq.~(\ref{KramersKronig}) that
\begin {equation} f_{xcL(T)}(0) \leq f_{xcL(T)}(\infty).
\label {bound1} \end {equation}
Recall now that the right hand side of Eq.~(\ref{bound1}) can be
expressed exactly, via the first moment of the
current-current spectral function,  in terms of
the expectation values of the kinetic and potential energy in the ground
state (a brief derivation is given in Appendix \ref{apphighfreq}):
\begin {equation}f_{xcL}(\infty) =
{1 \over  2n}\left[{12\over d} (\langle ke \rangle -
\langle ke \rangle_0) +{1+3\beta^{(d)} \over d}\langle pe \rangle\right]
\label {fxcLinfinity} \end {equation} and
\begin {equation}f_{xcT}(\infty) =
{1 \over  2n}\left[{4 \over d}(\langle ke \rangle -
  \langle ke \rangle_0) +{\beta^{(d)}-1 \over d}\langle pe
\rangle\right], \label {fxcTinfinity}
\end {equation}
where $\langle ke \rangle$, $\langle ke \rangle_0$, and $\langle pe
\rangle$ are the expectation values of the kinetic energy, the
noninteracting kinetic energy and the potential energy {\it
per particle} respectively, and $\beta^{(2)} = 1/2$, $\beta^{(3)}=1/5$.
These quantities can be expressed in terms of the exchange-correlation
energy per particle $\epsilon_{xc}(n)$ as follows:
\begin {equation}
\langle ke \rangle - \langle ke \rangle_0 = dn^{1/d+1}
\left ({\epsilon_{xc} \over n^{1/d}}\right )'
\end {equation}
and
\begin {equation}
\langle pe \rangle = - dn^{1+2/d}
\left ({\epsilon_{xc} \over n^{2/d}}\right )',
\end {equation}
where $d$ is the dimensionality and the prime denotes differentiation
of the function in the round brackets with respect to $n$.
The function $\epsilon_{xc}(n)$ is given in Refs.
\onlinecite{CeperleyAlder,Tanatar89} for $3$ and $2$ dimensions respectively.

The $\omega \to \infty$ limit of the {\it longitudinal}
local field factor was first calculated in three-dimensions
by Puff.\cite{Puff}  That result is usually referred to as the ``third
moment sum rule" \cite {SingwiTosi} since it is related to the third
moment of the dynamical structure factor -- the spectral function of the
density-density response  function $\chi$.
Because of  the relation $\chi = q^2 \chi_L/\omega^2$ (which follows from
gauge-invariance and from the continuity equation) the third moment of the
density-density response function coincides with the first moment of the
longitudinal current-current response function.
 The straightforward extensions
to two-dimensions and to the transverse case are outlined in Appendix
\ref{apphighfreq}.

Combining the foregoing  results with our expressions for the elastic
constants
[Eqs.~(\ref{mu}-\ref{K})] we obtain the rigorous inequalities
\begin{equation}
  \mu \leq \alpha_T^{(d)}nE_F +n^2 f_{xcT}(\infty)
  \label{boundmu}
\end {equation}
\begin{equation} \label{boundK}
  K + \left(2 - {2 \over d}\right) \mu \leq \alpha_L^{(d)}nE_F
  +n^2f_{xcL}(\infty),
\end {equation}
with the coefficients $\alpha$ defined after Eq.~(\ref{deffxc}).
The usefulness of these inequalities arises from the fact that the
quantities on the right hand sides are ground-state properties, which
can be calculated by Quantum Monte Carlo. Notice that the inequalities are
satisfied as strict equalities whenever the dissipation vanishes, i.e., when
$\Im f_{xcL(T)} =0$ at all frequencies.  In an electron liquid, this happens
both in the first order approximation with respect to the strength of the
Coulomb interaction  (weak coupling regime), and in the strong coupling
limit, when the electrons are expected to form a Wigner
crystal.\cite{Wigner34}   This observation leads us to suggest that
the right hand side
of  Eqs.~(\ref{boundmu}) and ~(\ref{boundK}) may provide  good approximation to
the elastic constants at {\it all} coupling strengths.

\section {Approximate expressions for the shear modulus and viscosity
of an electron liquid}
\label{secIV}
\subsection {High-density limit}
In the regime $na_B^d\gg 1$ where $a_B$ is the Bohr radius, the effect of the
Coulomb interaction is small, and can be treated by first-order
perturbation theory. It is straigthforward  to show that
the $f_{xcL(T)}(\omega)$'s are real and  independent of frequency in this
approximation.
This is because, in the limit $q \to 0$, the imaginary part of the
current-current response
functions
arises from processes involving at least {\it two}  electron-hole pair
excitations:  such processes are not allowed in first-order perturbation
theory.
The vanishing of ${\bf Im} f_{xcL(T)}(\omega)$, combined with the
dispersion relations ~(\ref{KramersKronig}) implies that ${\bf Re}
f_{xcL(T)}(\omega)$ is independent of frequency.  Hence Eq.~(\ref{bound1})
holds as a strict equality.
Since $\langle ke \rangle = \langle ke \rangle_0 = {d\over d+2} E_F$,
$\langle pe \rangle = -(3/4)e^2k_F/\pi$  for $d=3$  and   $\langle pe \rangle
= -(4/3)e^2k_F/\pi$  for $d=2$ in the first order approximation
we obtain
\begin {equation}
  \mu(n) = {2 n E_F \over 5} + {ne^2k_F  \over 10 \pi}
  \label {firstordermu3d} \end {equation}
(3 dimensions, $na_B^3\gg 1$) and
\begin {equation}
  \mu(n) = { n E_F \over 2} + {ne^2k_F  \over 6 \pi}
  \label {firstordermu2d} \end {equation}
(2 dimensions, $na_B^2\gg 1$). These results can also be obtained directly from
Eqs.~(\ref{mu},\ref{mu2d}) of Section III A, making use of the
first-order expression for the Landau parameters, $f_{\pvec \ppvec} = -
v(\pvec - \ppvec)/2$.

Let us now turn to the calculation of the high-density limit of the
viscosities $\eta$ and $\zeta$. Our starting point is the second-order
expression for
${\bf Im} f_{xcL(T)}(\omega)$, which is obtained from
Eq.~(11) of Ref.~\onlinecite{Nifosi} after replacing  the response functions
$\chi_{L,T}$ by the noninteracting ones $\chi^0_{L,T}$ and setting the
``exchange correction factor"  equal to $1$:

\begin {eqnarray} \label{NU1}
&&{\bf Im} f_{xcL(T)}(\omega) = - \int_0^\omega {d \omega ' \over \pi}
\int {d^d q \over (2 \pi)^d n^2} v(q)^2 \nonumber \\ &&\left[a_{L(T)}
  {q^2 \over
{\omega}^{'2}}
 {\bf Im} \chi^{(0)}_L(q,\omega ')  +b_{L(T)} {q^2
\over \omega^2}
{\bf Im} \chi^{(0)}_T(q,\omega ')\right] \nonumber \\ && {q^2 \over
(\omega - \omega
')^2} {\bf Im} \chi^{(0)}_L(q,\omega - \omega '),  \end {eqnarray}
with ($a_L$, $a_T$, $b_L$, $b_T$) equal to $(23/30,8/15,8/15,2/5)$ in
three dimensions and to $(11/16,9/16,1/2,1/2)$ in two dimensions.

The imaginary
parts of the noninteracting
response functions $\chi^{(0)}_{L(T)}$  at small $\omega$   and finite $q$
are directly calculated from  Eq.~(\ref{chi0s}):
\begin {equation} \label{chilsmallomega}
{q^2 \over \omega^2} {\bf Im} \chi^{(0)}_L(q,\omega) \simeq - {d \over 2}
{n \over E_F} \gamma_L {\omega \over q v_F} \end {equation}
and
\begin {equation} \label{chitsmallomega}
{\bf Im} \chi^{(0)}_T(q,\omega) \simeq - \gamma_T
{n \over m}  {\omega \over q v_F}, \end {equation}
for $\omega/v_F <q < 2q_F+\omega/v_F$;  zero otherwise.  The
constants $(\gamma_L, \gamma_T)$ are given by  $(\pi/2,3 \pi/4)$ for $d=3$ and
$(1,2)$ for $d=2$.

{}From the above formulas, it is easy to see that Eq.~(\ref{NU1}) gives
an infinite result,  due to the divergence of the unscreened Coulomb
interaction
$v(q)$ for $q \to 0$. The result is indeed finite if the screening
of the interaction is duly taken into account. In the high-density  limit
and at low frequency  this is accomplished  by the use of the Thomas-Fermi
statically screened interaction $v(q) \to v^{TF}(q)  =  4 \pi
e^2/(q^2+q_{TF}^2)$
($q_{TF}^2 = 6 \pi n e^2/E_F$) in three
dimensions, and $v(q) \to v^{TF}(q) =  2 \pi e^2/(q+q_{TF})$
($q_{TF}=2/a_B$) in
two dimensions.

The first term in Eq.~(\ref{NU1}), which involves the product of two
$\chi^{(0)}_L$'s,  is proportional to $\omega^3$, and therefore does not
contribute to the viscosity coefficients [see Eqs.~(\ref{eq3},
\ref{eq4})].  As for the second term, we find, after some
tedious but straightforward  calculations,
\begin {equation} \label {fxcsmallomega}
{\bf Im} f_{xcL(T)} = -\beta_{L(T)} {\omega \over E_F^2}
\int_0^{2q_F}q^{d-1} [v^{TF}(q)]^2 dq, \end {equation}
where $\beta_{L(T)} = (3/128 \pi)b_{L(T)}$ for $d=3$ and
$\beta_{L(T)} = (1/12 \pi^2)b_{L(T)}$ for $d=2$.

Evaluation of the integral and substitution in Eq.~(\ref{eq3}) leads to our
result for the high-density limit of the shear viscosity:

\begin{equation}\label{eqfiteta3d} \eta \simeq {n \sqrt\pi \over 40 (a_B
k_F)^{3/2}} \simeq {1\over 60} n r_s^{3/2} \end{equation} in three
dimensions, and  
\begin{equation} \label{eqfiteta2d} \eta \simeq  = {r_s^2 \over 6\pi} n \ln
{\sqrt2\over e r_s} = 0.053 r_s^2 n (-\ln r_s -0.65)  \end{equation}
in two dimensions.

It is interesting to notice that, in the same limit, the bulk viscosity $\zeta$
 vanishes, both in three and  two dimensions, due to
the relationship $b_L/b_T = 2(d-1)/d$.

\subsection {Low-density limit}
\label{seclowdensity}
In the regime  $na_B^d\ll 1$ the electron liquid is strongly correlated
(via the long-range Coulomb interaction) and its  behavior
is expected to be similar to that of a classical Wigner crystal.
\cite{Wigner34}  The elastic constants of a classical Wigner crystal
have been calculated by various authors.\cite{Maradudin60,Bonsall77} Of
particular interest is the case of the hexagonal lattice, which
is expected to be the stable crystal structure in {\it two dimensions}.
\cite{Bonsall77}  The elastic properties of this lattice
are formally
indistinguishable from those of a homogeneous and isotropic body, i.e.
there are only two elastic constants $K$ and $\mu$,\cite{Landau7} and
they are given by \cite{Bonsall77} $\mu \simeq 0.24 e^2 n^{3/2}$ and $K
= - 6 \mu$. (The fact that $K < 0$ in an electron liquid should be
no cause for alarm because this  bulk modulus  enters physical
properties summed to the Fourier transform of the Coulomb interaction
$v(q)$, which is large and positive at long
wavelength.) In three dimensions, the Wigner crystal has cubic symmetry, and
anisotropic elastic constants. The appropriate low-density limit for
the strongly
correlated liquid, obtained by averaging over different
orientations\cite{notaorientations}
is $\mu \simeq 0.19 e^2 n^{4/3}$ and $K = - (10/3) \mu$.

Remarkably, we find  that in this case, as well as in the weak coupling limit,
the inequality~(\ref{boundmu}) is obeyed as a strict equality, namely,
substituting on the right hand side of Eq.~(\ref{boundmu}) the
potential energy of the  Wigner crystal ($\sim -1.8/r_s Ryd$ in 3D,
$\sim -2.2/r_s Ryd$ in 2D), and
neglecting the kinetic energy, which tends to zero in the low density
limit, one obtains the correct value of $\mu$.  This implies that the
imaginary parts of $f_{xcL(T)}(\omega)$'s vanish in the low-density
limit.

\subsection {Interpolation formula}
At intermediate densities no exact results for the
$f_{xcL(T)}(\omega)$'s are available.  A mode-coupling calculation of
these quantities for both 2-dimensional and 3-dimensional electron
gases has recently been performed by Nifos\'\i\ {\em et al.}\cite{Nifosi}
Their results are expected to be an improvement upon previous
estimates,\cite {GrossKohn85} at least for not too large coupling
strengths.  Unfortunately, the values of $\mu$ obtained from this
approximate theory do not conform to the physical expectation that $\mu$
should reduce to the shear modulus of the classical Wigner crystal in the
limit of large $r_s$. In fact, the approximate $\mu$ is found to become
negative at  large $r_s$.  We believe that this should be
regarded as a failure of the approximate theory. For
this reason, we  propose an interpolation formula for $\mu$ that does not
suffer from this problem: it reduces to the correct limits for high and
low density, and does not differ substantially from the
estimates of Nifos\'\i\ {\em  et al.}\cite {Nifosi} in the range of $r_s$
where the latter are expected to be  reliable.

\begin{figure}
 \begin{center}
    \psfig{figure=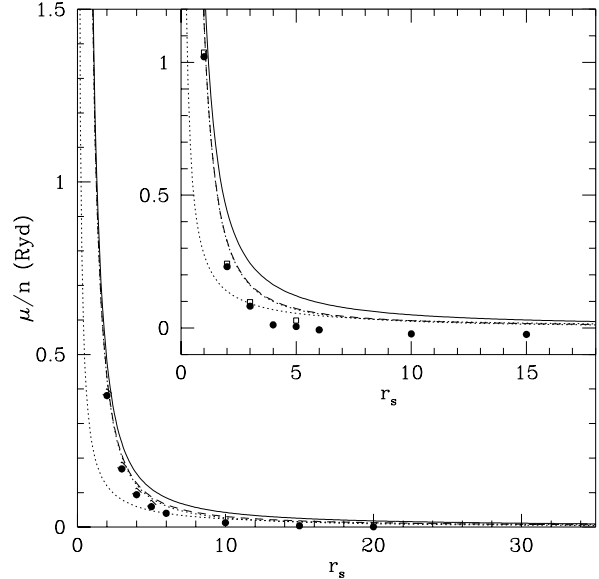,width=0.95\columnwidth}
  \end{center}
\caption{Shear modulus $\mu/n$ (full curve) as a function of
density in 3D (main figure) and 2D (inset). Dots are from
Ref.~\protect\onlinecite{Nifosi}, open squares have
been obtained from Eqs.~(\protect\ref{mu}-\protect\ref{mu2d}) using
estimates of
the Landau parameters from Refs.~\protect\onlinecite{Kwon94,Yasuhara92},
dotted curves are the
asymptotic behaviours of
Eqs.~(\protect\ref{firstordermu3d}-\protect\ref{firstordermu2d}) and from
the Wigner crystal (see Section \protect\ref{seclowdensity}), the full
curve is the upper bound of
Eq.~(\protect\ref{boundmu}), and the dashed curve is the approximate
interpolation of Eq.~(\protect\ref{muinterpolation}).}
  \label{figmu3d2d}
\end{figure}

Our approximate formula is
\begin {equation}
\mu/n =  a r_s^{-2} + b r_s^{-1} + (c-b) {1\over r_s+20}
\label {muinterpolation} \end {equation}
where $a$ and $b$ are obtained from the low-$r_s$ limit (in 3D,
$a= k_F^2/5m$ and $b={e^2 k_F/10\pi} $; in 2D,
$a=k_F^2/4m$ and $b=e^2 k_F/6\pi$)
and $c$ is obtained from the high-$r_s$ limit ($c=0.24 Ryd$ in 3D and
$c=0.22 Ryd$
in 2D). The approximate $\mu(r_s)$ is plotted in Fig.~\ref{figmu3d2d}, together
with  the values of
$\mu$ from Ref. \onlinecite{Nifosi}.

\begin{figure}
  \begin{center}
    \psfig{figure=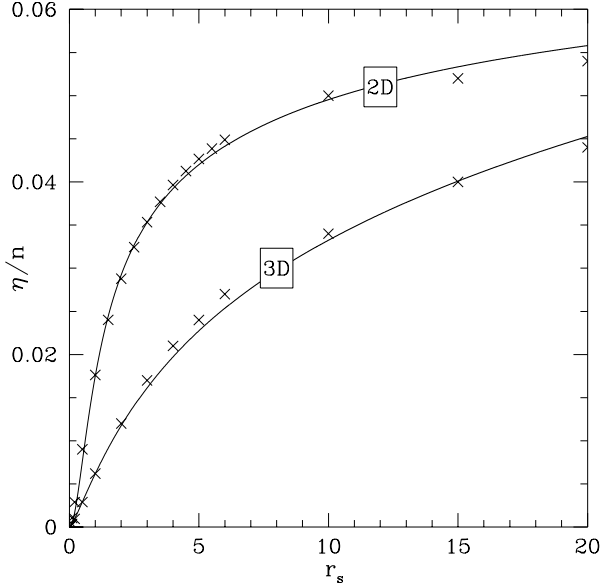,width=0.95\columnwidth}
  \end{center}
\caption{Shear viscosity $\eta$ in $d=2$ and $d=3$ (in units of $n$) from
the mode-coupling
calculation of Nifos\`\i\ {\em et al.}\cite{Nifosi} (crosses) compared with the
analytical expressions of Eqs.~(\protect\ref{eqfiteta3d}-\ref{eqfiteta2d})
(full curves).}
  \label{figeta}
\end{figure}

An analogous fit can be performed for the shear viscosity $\eta$, with the
{\em caveat} that in this case the high-$r_s$ behaviour is unknown and the
advantage that no sign problem is present in the mode-coupling computation.
The proposed formulas are
\begin{equation}
\eta \simeq \left(  60 r_s^{-3/2}
 + c_1 r_s^{-1} + c_2 r_s^{-2/3} + c_3
r_s^{-1/3}\right)^{-1} n
\end{equation}
where $c_1=80$, $c_2=-40$, $c_3=62$ in 3 dimensions, and
\begin{equation}
\eta \simeq \left[ \left({r_s^2\over 6\pi} \ln {\sqrt2\over e r_s} +
c_0 r_s^2 \right)^{-1} + c_1 r_s^{-2} + c_2 r_s^{-1/2} + c_3\right]^{-1} n
\end{equation}
where $c_0=0.25$, $c_1=21$, $c_2=23$, $c_3=13$.
in 2 dimensions. The resulting functions are compared in Fig.~\ref{figeta} with
the  values calculated in Ref. \onlinecite{Nifosi}.

\subsection {Transverse collective mode in the electron liquid}

A transverse collisionless sound mode will exist in the uniform
electron liquid provided that the transverse current-current response function
has a pole close to the real axis.  Unfortunately, because such a mode
must have a linear dispersion of the form $\omega = c_t q$
 the visco-elastic
form  of the response function [Eq.~(\ref
{viscoelasticresponset})] - which is valid for $\omega \gg qv_F$ - is not
 applicable unless $c_t \gg v_F$.  One should instead use the response function
in the limit $\omega/q =c_t = constant$,  which involves all the Landau
parameters, and is presently unknown.

The situation becomes much more favorable in the limit of large $r_s$
(low density). In this limit, the transverse sound velocity
{\it is} large compared to the Fermi velocity, so that the visco-elastic
form  of the transverse response function
 can be used.  From Eq.~(\ref {viscoelasticresponset}) one can
immediately deduce the existence of a pole at
\begin {equation}
\omega = c_t q - i {\eta \over 2 mn} q^2,
\label {transversedispersion} \end {equation}
where $c_t^2 = \mu/mn$.  This result is independent of dimensionality.
Note that the linewidth of the excitation ($\eta q^2 /nm$)  vanishes in
the long wavelength limit.

{}From now on let us focus on the two-dimensional electron liquid, since
this is the system which, being closer to Wigner crystallization, provides
the best chances for the observation of a transverse sound mode.
At low density, making use of  Eq.~(\ref{transversedispersion}) and
of  the low-density form the shear modulus, we obtain \begin {equation}
{c_t^2 \over v_F^2}  \simeq 0.07 r_s,
\label {ctratio} \end {equation}
which  grows with decreasing density, and justifies the use of the
visco-elastic form of the response function.

Equation ~(\ref{ctratio}) can be used to give a rough estimate of the
minimum value of $r_s$ above which the transverse sound mode would be
observable.  Requiring $c_t/v_F> 1$ we obtain the condition $r_s > 14$.  This
minimum value of $r_s$ is still significantly lower than the critical $r_s
\simeq37$ for Wigner crystallization, as estimated from Monte Carlo
calculations.\cite{Tanatar89}

\section {Inclusion of collisions}
\label{secV}
\subsection {The relaxation time approximation}
Up to this point, we have neglected  processes that
limit the lifetime of a quasiparticle, such as
quasiparticle-impurity and  quasiparticle-quasiparticle collisions.

In this Section we  reinstate these processes, and  study
their effect at the phenomenological level.  The starting point of our
analysis is still the kinetic equation for the quasiparticle
distribution function, but now we include a  collision term:
\begin {eqnarray}
  &&(\vec q \cdot \vec v_p - \omega) n_{1p}(\vec q,\omega) -\vec q \cdot
  \vec v_p n_0'(\epsilon_p) \nonumber\\
  &&\times\left[ \sum_{p'}f_{pp'}n_{1p'}(\vec q,\omega) +
    {\vec p \over m}\cdot \vec A(\vec q,\omega)\right] =-iI[ n_{1p}],
  \label {collisiontransportequation}
\end {eqnarray}
where $I[n_{1p}]$ is the collision integral.

Without going into the details of the collision process, we shall
simply assume that collisions attempt to restore a ``locally relaxed"
equilibrium distribution function $n_p^\rel \equiv n_0 (\epsilon_p) +
n_{1p}^\rel$, with a characteristic relaxation time $\tau$, namely
\begin {equation}
  I[ n_{1p}]  = -{n_{1p} -n_{1p}^\rel \over \tau}\,.
  \label{collisionintegral}
\end {equation}

Equation (\ref{collisionintegral}) is generally referred to as
the ``relaxation time approximation".\cite{Mermin70Das75}

The locally relaxed distribution function $n_p^\rel (\vec r,t)$ is defined
as the
distribution that, at any given instant,   would be in
{\it equilibrium} in the presence of appropriate scalar  and vector potentials
$V_\rel(\vec r,t)$ and $\Avec_\rel (\vec r,t)$, chosen so as to make
Eq.~(\ref{collisiontransportequation}) obey the conservation of particle number
and (when appropriate) particle current. Let us discuss this construction in
some detail.

{\it (i) Impurity scattering.} In this case collisions conserve the
local quasiparticle number (i.e., the density), but not the current.
Therefore, we must require
\begin {equation}
  \sum_p I [n_{1p}] \sim \sum_p(n_{1p}-n_{1p}^\rel) =0,
  \label{numberconservation}
\end{equation}
that is, the locally relaxed   distribution function must yield the same
density
as the true distribution function
\begin {equation} n_1(q,\omega) \equiv \sum_p n_{1p}(q,\omega) =
\sum_p n_{1p}^\rel (q,\omega). \end {equation}
This is accomplished by defining $n_p^\rel$ as the instantaneous
equilibrium
distribution function in the presence of a scalar potential $V_\rel$ such
that \begin {equation} \label{vrel}  V_\rel  = \chi^{-1} (0) n_1(q, \omega)
\end
{equation}
 where $\chi(0)$ is the static ($\omega =0$) density-density response function
in the $q \to 0$ limit.
Because there are no additional constraints,
the vector potential $\Avec_\rel$  remains  equal to  the real one
$\vec A$.

{\it (ii) Quasiparticle-quasiparticle scattering.} In this case, the
collisions conserve not only the local number, but also the local
momentum, i.e., the current density.  Therefore, in addition to
(\ref{numberconservation}),  we must require
\begin {equation}
  \sum_p \vec p \, I [\delta n_p] \sim \sum_p \vec p
  (\delta n_p-\delta n_p^\rel)=0,
  \label{currentconservation}
\end{equation}
that is, the locally relaxed   distribution function must yield the same
canonical current density as the true distribution function:
\begin {equation} \vec j_{c}(q,\omega) \equiv \sum_p n_{1p}(q,\omega) \vec
p/m =
\sum_p n_{1p}^\rel (q,\omega) \vec p/m. \end {equation}
On the other hand, the {\it full} locally relaxed current density
\begin {equation}
\vec j^\rel = \vec j_c + n \vec A_\rel /m \end {equation}
must vanish, because it is the current of a system in {\it equilibrium}. This
fixes the value of the $\vec A_\rel$  potential as
\begin {equation} \label{arel}
 \vec A_R (q,\omega) = - (m/n)
j_c(q,\omega). \end {equation}  The value of $V_\rel$ is still given
by Eq.~(\ref{vrel}).

\subsection {Solutions of the transport equation in the relaxation
time approximation}
Equation ~(\ref{collisiontransportequation}) can be solved  to yield the
density-density ($\chi$) and the transverse current-current ($\chi_T$) response
functions of the system {\it with}  collisions, in terms of those of the system
{\it without} collisions, i.e., with $1 /\tau$ set  to zero. We describe our
method of solution in Appendix \ref{appsollandRTA}.

In the case of {\it impurity scattering} (no current conservation) we
obtain \begin {equation}
{1 \over \chi^\tau(\vec q, \omega)} = {\omega \over \omega+i/\tau}
{1 \over \chi(\vec q, \omega + i/\tau)} +{i /\tau \over \omega +
i/\tau} {1 \over \chi(q,0)}, \label{chitau} \end {equation}
where $\chi^\tau (\vec q,\omega)$ is the density-density response
function including collisions, and $\chi(q, \omega+i/\tau)$ is the same
quantity without collisions, but calculated at the complex frequency
$\omega +i/\tau$.
Similarly, for the transverse current-current response function we
obtain:
\begin {equation}
{1 \over \chi_T^\tau(\vec q, \omega)} = {\omega+i/\tau \over \omega}
{1 \over \chi_T(\vec q, \omega + i/\tau)}. \label{chiTtau} \end
{equation}
The longitudinal current-current response function is, of course, obtained
from the density-density response function via the continuity equation
relation $\chi_L^\tau(q,\omega) =
 (\omega^2/q^2) \chi^\tau (\vec q,\omega)$, which continues to hold in
the presence of scattering. We note, in passing, that the above
equations, when used to calculate the electrical conductivity, lead to
the familiar Drude formula,  where $\tau$ is the electron-impurity
scattering time.

In the case of {\it current-conserving scattering} the above two equations are
modified as follows:
\begin {eqnarray}
{1 \over \chi^\tau(\vec q, \omega)} &=& {\omega \over \omega+i/\tau}
{1 \over \chi(\vec q, \omega + i/\tau)} \nonumber\\
&&+{i /\tau \over \omega +
i/\tau} {1 \over \chi(q,0)} - {i \over \tau}{m \over n} {\omega \over
q^2} \label{chitau2} \end {eqnarray}
and
\begin {equation}
{1 \over \chi_T^\tau(\vec q, \omega)} = {\omega +i/\tau\over \omega}
{1 \over \chi_T(\vec q, \omega + i/\tau)}-{i \over \omega \tau} {m \over
n}. \label{chiTtau2} \end {equation}
Notice the additional terms on the right hand sides of these equations,
which guarantee current conservation.

In both cases, we define the  exchange correlation kernels
$f_{xcL(T)}^\tau (\omega)$ in the presence of collision, by direct
generalization of Eq.~(\ref{deffxcnew}), namely
\begin {equation}
f_{xcL(T)}^\tau(\omega) = \lim_{q \to 0} {\omega^2 \over q^2} \left[{1 \over
\chi_{L(T)}^{0 \tau}(\vec q, \omega)} -
{1 \over
\chi_{L(T)}^{\tau}(\vec q, \omega)}\right],
 \label{deffxctau} \end {equation} where the ``reference" response
function  $\chi_{L(T)}^{0 \tau}(\vec q, \omega)$ in the presence of
collisions, is obtained from the solution of the kinetic equation
~(\ref{collisiontransportequation}) with all the Landau parameters set
equal to zero.

We notice that our ``reference function'' is not  the same as the
noninteracting response function,  because it contains the relaxation
time which is  determined, at least in part, by
electron-electron interactions. Thus
$f_{xc}^\tau$ is  a mathematical construct: its purpose is to take into account
{\it some}  interaction effects which admit description in terms of
Landau parameters. Additional interaction effects are phenomenologically
included
in the relaxation time, and are already contained in the reference response
function.

With the above definitions, the collisional exchange-correlation kernels
are found to be related to the collisionless kernels by  relationships that
closely parallel the analogous ones for the inverse response functions:
\begin {equation}
f_{xcL}^\tau(\omega) = {\omega \over \omega+i/\tau}
f_{xcL}( \omega + i/\tau) +{i /\tau \over \omega +
i/\tau} {d^2 \epsilon_{xc}(n) \over dn^2} \label{fxcLtau} \end {equation}
and
\begin {equation}
f_{xcT}^\tau(\omega) = {\omega+i/\tau \over \omega}
f_{xcT}( \omega + i/\tau) \label{fxcTtau} \end {equation}
Note that these formulas hold for both types of scattering.

In practice, under the assumption that $1 /\tau \ll  E_F$, one can
approximate $f_{xcL(T)}(\omega+ i /\tau) \simeq f_{xcL(T)}(\omega)$.
This is justified because the {\it collisionless} $f_{xc}$'s are
smooth functions of $\omega$, which vary significantly on a scale set by
the  Fermi energy  (or plasmon frequency).
Therefore, the fractional error introduced  by neglecting
$1/\tau$ in the argument of $f_{xc}$'s is expected to be of  order $1/E_F
\tau \ll 1$.

Equations~(\ref{fxcLtau}) and (\ref{fxcTtau}) provide the basis for
an approximation to the frequency-dependence of the
exchange-correlation kernels, which interpolates smoothly between the
static limit and the dynamic low-frequency limit across a region of
width $1 /\tau$ in frequency.  The form of the dependence of $f_{xc}$ on
the inverse scattering times shows that the following additivity
property holds:  If there are two independent scattering mechanisms
operating simultaneously with relaxation times $\tau_1$ and $\tau_2$,
then their combined effect is equivalent to that of a single scattering
mechanism with an effective relaxation time $\tau_{eff}$ such that
\begin {equation}
  {1 \over \tau_{eff}} = {1 \over \tau_1}+{1 \over \tau_2}\,.
  \label{additivity}
\end{equation}
This can be proved straightforwardly, by
applying the transformations  ~(\ref{fxcLtau}) and ~(\ref{fxcTtau}) twice
in succession, the first time with relaxation time $\tau_1$ and the second
time with relaxation time $\tau_2$.  The result is the same that one would
obtain by applying the transformation only once, with relaxation time
$\tau_{eff}$.

\subsection  {Elastic constants and viscosity in the presence of
collisions}

In order to calculate the elastic constants and viscous coefficients in
the presence of collisions  we first substitute in Eqs. ~(\ref{chitau})
and ~(\ref{chiTtau}) the long-wavelength forms of the {\it collisionless}
response functions derived in Section II.  These are conveniently
rewritten as
\begin {equation}
\chi (\vec q, \omega) \sim {nq^2 \over m \omega^2} \left\{1 +\left[ {K
      \over n^2} +\left(2 - {2 \over d}\right) {\tilde \mu (\omega)
      \over n^2} \right] {nq^2 \over m \omega^2}\right\}
\label {chilongwave} \end {equation}  and
\begin {equation}
\chi_T (\vec q, \omega) \sim {n \over m} \left[1 + {\tilde \mu (\omega) \over
n^2} {nq^2 \over m \omega^2}\right] \label {chiTlongwave}, \end {equation}
where $\tilde \mu(\omega)$ is the collisionless generalized shear
modulus, and where we have taken into account the fact that, according to our
previous discussion, the bulk viscosity of the electron gas
vanishes, and  $\tilde K(\omega) \simeq K$, at low
frequency.  We also need the long-wavelength form of the static density-density
response function, which is
\begin {equation}
\chi(\vec q,0) \sim - {n^2 \over K}. \label {chi0longwave} \end {equation}
Then, Eqs.~(\ref{chitau}) and (\ref{chiTtau}) yield, in the case of
impurity scattering,
\begin {eqnarray}
{1 \over \chi^\tau(\vec q, \omega)} &=& {m  \omega(\omega+i/\tau)\over
nq^2} - {\omega \over \omega+i/\tau}
\left[{K \over n^2}\right.
 \nonumber\\&& \left. +\left(2 - {2 \over d}\right) {\tilde \mu (\omega
    +i/\tau) \over n^2}\right]
-{i/\tau \over \omega+i/\tau} {K \over n^2} \label{chitaulongwave}
\end{eqnarray}
and  \begin{equation}
{1 \over \chi_T^\tau(\vec q, \omega)} = {m  (\omega+i/\tau)\over
n \omega}   - {\omega \over \omega+i/\tau}
 {\tilde \mu (\omega +i/\tau) \over n^2}{q^2 \over
\omega^2}.\label{chiTtaulongwave} \end {equation} Again, we can neglect
$i/\tau$ in the argument of $\tilde \mu$ with a relative error of order
$1/E_F\tau$.

Next, we compare the long wavelength forms of the  response
functions ~(\ref{chitaulongwave},\ref{chiTtaulongwave}) with the ones obtained
from generalized elasticity theory  in the presence of collisions.
In order to do this,  we return to
Eq.~(\ref{viscoelasticmotion}) and add a relaxation term $-\vec j_1
(\vec q, \omega) /\tau$ on the right hand side in the case of impurity
scattering  (in the case of current-conserving scattering no additional term is
needed).  The  elastic constants must also be modified to include the
effect of  collisions: we call $\tilde K^\tau$ and $\tilde \mu^\tau$ the
generalized elastic constants, which depend on frequency on the scale of
$1/\tau$.  Solving the modified equations of motion
 we obtain the   density-density and transverse current-current response
functions in the following form
\begin {equation} {1 \over \chi^\tau(\vec q, \omega)} = {m
\omega(\omega+i/\tau)\over nq^2}   -  \left[{\tilde K^\tau \over n^2}
+\left(2 -
{2 \over d}\right) {\tilde
    \mu^\tau (\omega)
\over n^2}\right]  \label{chitaumacro} \end {equation}
and \begin{equation}
{1 \over \chi_T^\tau(\vec q, \omega)} = {m  (\omega+i/\tau)\over
n \omega}   -
 {\tilde \mu^\tau (\omega) \over n^2}{q^2 \over
\omega^2}.\label{chiTtaumacro} \end {equation}
Comparing equations~(\ref{chitaumacro}) and~(\ref{chiTtaumacro}) to
 ~(\ref{chitaulongwave}) and~(\ref{chiTtaulongwave}) respectively we
accomplish our  goal of expressing the  elastic
constants $\tilde K^\tau$ and $\tilde \mu^\tau$  in
terms of their collisionless counterparts $\tilde K$ and $\tilde \mu$:
\begin {equation}
\tilde \mu^\tau (\omega) = {\omega \over \omega +i/\tau} \tilde
\mu(\omega) \label{tildemutau} \end {equation}  and
\begin {equation} \tilde K^\tau (\omega) = \tilde K(\omega) =K.
\label{tildeKtau} \end {equation}

It is straigthforward to verify that the same results are also obtained
in the case of current-conserving scattering.

{}From Eq.~(\ref{tildeKtau}) we see that the bulk modulus and the bulk
viscosity coefficient are unaffected by collisions.  For the
shear modulus and the shear viscosity  we obtain, after separating
the real and the imaginary parts of Eq.~(\ref{tildemutau}),
the following equations, accurate  within corrections of order $1/E_F
\tau$:
\begin {equation}
\mu^\tau = \mu {(\omega \tau)^2 \over 1 +(\omega \tau)^2 }, \label
{mutau} \end {equation} and
\begin {equation} \eta^\tau = {\mu  \tau} {1 \over
1 +(\omega \tau)^2} + \eta {(\omega \tau)^2 \over 1 +(\omega \tau)^2}.
\label{etatau} \end {equation}

In reaching the final form of these equations, we have used the fact
that $\eta/\tau \sim 1/E_F \tau \ll  \mu$ where $\eta$ and $\mu$ are the
{\it collisionless} viscosity and shear modulus.

\begin{figure}
  \begin{center}
    \psfig{figure=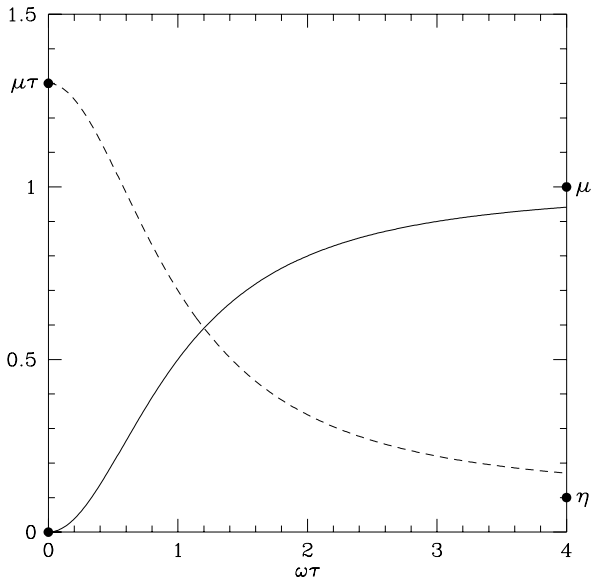,width=0.95\columnwidth}
  \end{center}
\caption{Shear modulus $\mu^\tau$ (full curve) and viscosity
  $\eta^\tau$ (dashed curve) in the presence of collisions, as
  functions of $\omega\tau$, from Eqs.~(\ref{mutau}-\ref{etatau}). The
  dots mark the $\omega\tau=0$ and $\omega\tau=\infty$ asymptotic limits.}
  \label{figsh}
\end{figure}

Equations ~(\ref{mutau}) and ~(\ref{etatau}) clearly exhibit the
crossover from hydrodynamic to dynamic regime.  Their qualitative behavior is
plotted in Figure \ref{figsh}, and one can observe the
opposite behaviors
of $\eta^\tau$ and $\mu^\tau$  as functions of frequency, which is
consistent with the Kramers-Kr\"onig dispersion relations.

\section{Conclusion}
In this paper we have shown that visco-elasticity is the effective
theory that describes the dynamical response of a Fermi liquid at low
temperature, long wavelength and low (but finite) frequency.
We have presented rigorous results and approximate expressions for the
visco-elastic coefficients of an electron liquid in two and three
dimensions.   We have
also shown how quasiparticle scattering mechanisms can be incorporated
in the effective theory by means of a simple relaxation time
approximation.  An interesting prediction of our work is the possibility of
the existence of a transverse sound mode in a high purity
two-dimensional electron liquid at large $r_s$, before
crystallization occurs.

\section{Acknowledgements}
This work was supported by NSF grant No. DMR-9706788.
Sergio Conti acknowledges a travel grant from the
Max-Planck-Institute for Mathematics in the Sciences.
Giovanni Vignale was also supported in part by the Institute
for  Theoretical Physics at the University of California,
Santa Barbara, under NSF Grant No. PHY94-07194.
 \label{secconcl}

\appendix

\section{High-frequency limits of the exchange-correlation kernels}
\label{apphighfreq}
This Appendix gives a self-contained derivation of the high-frequency
limits of the exchange-correlation kernels, based on the equation of
motion for the current-current response function.

The current-current response function $\chi_{ij}(\qvec,\omega) =
(n/m)\delta_{ij}+ R_{ij}(\qvec,\omega)$ is determined by the Fourier transform
of\cite{SingwiTosi} \begin{equation}
R_{ij}(\qvec,t) = -i\theta(t) \langle [
j_{\qvec,i}(t),j_{-\qvec,j}(0)]\rangle
\end{equation}
where $\jvec_\qvec = \sum_\pvec (\pvec/m) c^\dagger_{\pvec-\qvec/2}
c^{}_{\pvec+\qvec/2}$ is the canonical  current operator.
The time derivative of $R_{ij} (\qvec,t)$ evaluated at $t=0$ gives the first
frequency moment of $R_{ij}(\qvec,\omega)$,
\begin{equation}\label{eqappmij}
M_{ij}(\qvec) \equiv i \int_{-\infty}^\infty \R_{ij}(\qvec,\omega) \omega
{d\omega \over \pi}  = \left. {d\over
dt}  \langle [
j_{\qvec,i}(t),j_{-\qvec,j}(0)]\rangle \right|_{t=0}
\end{equation}
which is clearly real, i.e. only the imaginary part of
$\R_{ij}(\qvec,\omega)$ contributes to the frequency integration.
We now evaluate the right hand side of Eq.~(\ref{eqappmij}). This gives
\begin{equation}
 \left. {d\over
dt}  \langle [
j_{\qvec,i}(t),j_{-\qvec,j}(0)]\rangle \right|_{t=0} =
 \langle [ [H,
j_{\qvec,i}],j_{-\qvec,j}]\rangle
\end{equation}
where $H$ is the Hamiltonian and all operators on the right hand side are
evaluated at $t=0$. Straightforward evaluation of the commutators  allows us to
obtain $M_{ij}(\qvec)$ in terms of the momentum distribution
$n_\pvec=\langle c^\dagger_\pvec c^{}_\pvec\rangle$ and the static
structure factor $S(q)$ of the electron liquid. The resulting
expression is, up to terms of order $q^2$
\begin{eqnarray}
 M_{ij}(\qvec) &=& \sum_\pvec n_\pvec
{q^2 p^ip^j +(\pvec\cdot\qvec)(p^iq^j + q^i
p^j)\over m^3}+\nonumber\\
&&+  v(p) S(\pvec) \left[ {q^iq^j\over m^2} + (1-d)
{\qvec\cdot\pvec \over p^2} {p^iq^j+p^jq^i\over m^2}\right.\nonumber\\
&&\hskip1cm\left. + {p^ip^j\over
m^2} \left( {1-d\over2} {q^2\over p^2} + {d^2-1\over2}
{(\pvec\cdot\qvec)^2\over p^4} \right)\right],
\label{eqfxcinftyij}\end{eqnarray}
where  inversion symmetry has been used to eliminate terms of first
order in $q$.

{}From the
Kramers-Kronig relations one obtains
the high-frequency behaviour of the real part of $\chi_{ij}(\qvec,\omega)$,
\begin{equation}
\Re\, \chi_{ij}(\qvec,\omega\to\infty) \simeq {n \over m} \delta_{ij}
+{1\over \omega^2} M_{ij}(\qvec)\,. \end{equation}
The longitudinal and transverse components of $\chi$ are then obtained from
$\chi_L = \hat q_i \chi_{ij} \hat q_j$ and
$\chi_T = \hat t_i \chi_{ij} \hat t_j$.  ($\hat t$ is a unit vector
perpendicular to $\hat q = \qvec/q$; a sum over repeated indices is
implied).  Comparison with Eq.~(\ref{chismallq}) yields
\begin{equation}
\lim_{\omega\to\infty} f_{xc L(T)}(\omega) = \lim_{\qvec\to0} {m^2
\over n^2 q^2} M_{L(T)}(\qvec)\, -{\alpha_{L(T)} E_F \over n}.
\end{equation}

In an isotropic system, the first line in Eq.~(\ref {eqfxcinftyij}) is
proportional to the average kinetic energy, and the remaining to the
average potential energy, leading to
Eqs.~(\ref{fxcLinfinity}-\ref{fxcTinfinity}) of the main text.
We finally remark that full isotropy is not needed to obtain
Eqs.~(\ref{fxcLinfinity}-\ref{fxcTinfinity}) -- indeed, only averages
of second- and fourth-order terms appear in Eq.~(\ref{eqfxcinftyij}).
In the case of the 2D triangular lattice, such averages are identical
to those of an isotropic fluid ($\langle x^2\rangle=r^2/2$, $\langle
x^4\rangle = 3r^4/8$, $\langle x^2 y^2\rangle=r^4/8$), and therefore
the same results hold in the crystal and in the liquid. This is not
the case in any of the crystals with cubic symmetry, such as simple
cubic in 2D and fcc in 3D.

\section{Solution of Landau equation of motion within the RTA}
\label{appsollandRTA}

This Appendix discusses the details of the computation of the response
functions within the RTA. As in the main text, we denote by
$\chi^\tau(\qvec,\omega)$ the response function in the presence of a
relaxation time $\tau$, and by $\chi(\qvec,\omega+i/\tau)$ the
collisionless response function evalutated at the complex frequency
$\omega+i/\tau$. The dependence on $\qvec$, which plays no significant
role in the computation, will  henceforth be dropped.

The collision integral ~(\ref{collisionintegral}) describes
relaxation to a  local equilibrium distribution $ n_p^\rel$,
which was defined as the equilibrium solution in the presence of
appropriate vector and scalar potentials $\Avec_\rel$ and $V_\rel$.

In Section V we determined the
value of $\Avec_\rel$ and $V_\rel$ from  the condition that the transport
equation ~(\ref{collisiontransportequation}) obey the conservation of
particle number and (when appropriate) particle current. The results of those
calculations will be crucial in the following development.

In order to facilitate the solution of the transport equation, it is convenient
to introduce a  ``dynamic correction to the quasiparticle distribution
function", defined as follows   \begin{equation}
  n_{1p}^\dyn \equiv   n_{1p} - {i/\tau\over \omega + i/\tau}
 n_{1p}^\rel\,.
\end{equation}
Making use of the equilibrium condition for $n_{1p}^\rel$  it
is a straightforward computation to verify that the
collisional transport equation (\ref{collisiontransportequation})
for $n_{1p}$ is equivalent to
the {\em collisionless} transport equation for $n_{1p}^\dyn $ with
a modified frequency
$\omega+i/\tau$ and modified potentials
\begin{equation}\label{eqdefa3}
  \Avec_\dyn =\Avec  - {i/\tau\over \omega + i/\tau}
  \Avec_\rel
\end{equation}
and
\begin{equation}
  V_\dyn =V  - {i/\tau\over \omega + i/\tau}
  V_\rel \,.
\end{equation}
It follows that the density and current responses
$n_1^\dyn = \sum_p n_{1p}^\dyn
$,  $\vec j^\dyn = \sum_p n_{1p}^\dyn \vec p /m + n_1^\dyn \vec A^\dyn /m$
 are related to the modified fields $\Avec_\dyn $, $V_\dyn $  by the
collisionless response
functions evaluated at frequency $\omega+i/\tau$,
\begin{equation}\label{eqrespdens3}
  n_1^{(\dyn )}(\qvec,\omega)=
  \chi(\omega+i/\tau) \left[V_\dyn
    + {(\omega+i/\tau) \qvec\cdot \Avec_\dyn \over q^2 }\right]
\end{equation}
and
\begin{equation} \label{defjd}
  \jvec^{(\dyn)} = \chi_{L,T}(\omega+i/\tau)
  \Avec_\dyn ^{L,T} + \chi_L(\omega+i/\tau) {\qvec\over\omega+i/\tau} V_\dyn
\end{equation}

In writing these equations we have used the fact that the mixed density-current
response functions $\vec \chi_{dc}$ and $\chi_{cd}$ are purely longitudinal,
and are related to the density-density response  function by $\vec \chi_{cd}  =
\vec \chi_{dc} = \vec q \omega \chi /q^2 = \vec q \chi_L /\omega$.

Finally, note that $n_1^{(\dyn)} (\omega)= n_1(\omega) - {i/\tau\over\omega
  +i/\tau} n_1^{(\rel)}(\omega)$, and that a similar relation $
\vec j_c^\dyn(\omega) = \vec j_c(\omega) - {i/\tau \over \omega + i/\tau} \vec
j_c^\rel(\omega)$  holds between the
  canonical currents $\jvec_{c}^\dyn$ and $\jvec_{c}^\rel$ as well as  between
the full currents  $\jvec^\dyn$ and $\jvec^\rel$. Since
  $\jvec^{\rel}=0$ (the full current vanishes in an equilibrium
state) one concludes that $\jvec=\jvec^{(\dyn)}$.  We now discuss the
two different cases mentioned in Section \ref{secV}.

\subsection{Impurity scattering}
In this case (see Section V) $\Avec_\rel =\Avec_1$ and $V_\rel $ is fixed from
the constraint of local density conservation,
$n_1^{(\rel)}(\qvec,\omega)=n_1(\qvec,\omega)$.

\paragraph{Density excitations, scalar external potential}
In this case both $\vec A$ and $\vec A^\rel$  vanish. Density conservation
gives
[from Eq.~(\ref{vrel})]
\begin{equation}\label{eqv2densimp}
  V_\rel= {n_1 \over \chi(0)} =  V_1 {\chi^\tau(\omega)\over \chi(0)}\,.
\end{equation}
Since $n_1^{(\rel )}=n_1$, we get
\begin{equation}
n_1 =
 {\omega +i/\tau\over \omega}
n_1^{(\dyn )}=
 {\omega +i/\tau\over \omega}
 \chi(\omega+i/\tau) V_\dyn
\end{equation}
i.e.
\begin{equation}
  \chi^\tau(\omega) =  {\omega +i/\tau\over \omega}
\chi(\omega+i/\tau) \left[1 - {\chi^\tau(\omega)\over \chi(0)} {i/\tau\over
    \omega+i/\tau}\right]
\end{equation}
which, after some algebra, gives Eq.~(\ref{chitau}).

\paragraph{Density excitations, longitudinal vector potential}
In this case the scalar external potential $V=0$ and there is a purely
longitudinal external vector potential $\vec A$.  This is completely equivalent
to the previous case, modulo a gauge transformation. We carry out the
computation only as a check. The total density fluctuation is \begin{equation}
  n_1(\qvec,\omega) = \chi_L^\tau(\omega) {\qvec\cdot \vec A (\qvec,\omega)
  \over \omega}
\end{equation}
and the potentials of the fictitious system are given by
$\Avec_\rel =\Avec$, $V_\rel  = [\chi_L^\tau(\omega)/\chi(0)] \qvec\cdot
\Avec/\omega$.
If follows that
\begin{equation}
  \jvec = \chi_L(\omega+i/\tau) \left[  {\omega\over
  \omega+i/\tau} - {i/\tau\over (\omega + i/\tau)^2} {q^2\over \omega}
  {\chi_L^\tau(\omega)\over\chi(0)} \right] \Avec
\end{equation}
which gives
\begin{equation}
  {\omega (\omega+i/\tau)\over q^2} {1\over \chi_L(\omega + i/\tau)} =
{\omega^2\over q^2}  {1\over \chi_L^\tau(\omega)} - {i/\tau\over \omega +
  i/\tau} {1\over\chi(0)}
\end{equation}
Since $\chi_L(\omega+i/\tau) = (\omega+i/\tau)^2 q^{-2}
\chi(\omega+i/\tau)$ this is equivalent to the previous result of
Eq.~(\ref{chitau}).

\paragraph{Transverse excitations}
The external vector potential $\vec A$ is purely transverse. There is no
density
fluctuation,  $V_1=V_\rel =V_\dyn =0$, and $\Avec_\rel =\Avec_1$. Thus, from
the combination of Eqs.~(\ref{eqdefa3}) and ~(\ref{defjd}) we obtain
\begin{equation}
  \jvec(\qvec,\omega) =
  {\omega\over\omega+i/\tau} \chi_{T}(\omega+i/\tau)
  \Avec(\qvec,\omega)
\end{equation}
which is equivalent to Eq.~(\ref{chiTtau}).

\subsection{Current-conserving scattering}
Here $V_\rel $ is fixed as in Eq.~(\ref{eqv2densimp}) , and moreover
$A_\rel $ is
given by  Eq.~(\ref{arel}),  $\Avec_\rel =-(m/n) \jvec_c$.

\paragraph{Density excitations, scalar external potential}
Here $\Avec=0$ and $\vec j_c = \vec j$ (the full current and the canonical
current coincide).  Thus  $\Avec_\rel =-(m/n) \jvec = -(m\omega/nq) \chi^\tau
V_1$, and $V_\rel $ is the same as in Eq.~(\ref{eqv2densimp}). It follows that
\begin{eqnarray}
  n_1^{(\dyn)} &=&  {\omega\over\omega+i/\tau} n_1 \\
&=& \chi(\omega+i/\tau) \left[ 1 -
  {\chi^\tau(\omega)\over\chi(0)} {i/\tau\over \omega + i/\tau} + {i\over
    \tau} {m\omega\over n q^2} \chi^\tau(\omega)\right] V\nonumber
\end{eqnarray}
which is equivalent to Eq.~(\ref{chitau2}). The same computation can
be done using a longitudinal vector potential, with the same result.

\paragraph{Transverse excitations}
Here $V=V_\rel =0$; and $\Avec_\rel  = - (m/n) \jvec_{c} = (1 -
m \chi_T^\tau(\omega)/n) \Avec$, which gives
\begin{equation}
  \Avec_\dyn  = \left[{\omega\over \omega+i/\tau} +
  {i/\tau\over \omega+i/\tau}
{m\over n} \chi_T^\tau(\omega)
 \right] \Avec
\end{equation}
Finally, since
$  \chi_T^\tau(\omega)\Avec = \chi_T(\omega+i/\tau)\Avec_\dyn $
we get Eq.~(\ref{chiTtau2}).


\begin{thebibliography}{10}
\bibitem {Footnote0} Such is, for instance, a polycrystalline solid.
\bibitem {Footnote00} The tranverse part of the vector potential gives
rise to a magnetic field that exerts a Lorentz force on the electric
current. It is evident that the Lorentz force is a second order effect,
since the induced current is a first order effect: it therefore does not
play any role in linear response theory.
\bibitem{notaorientations} The energy of a deformation field $u_i(\qvec)$
is in general given by $E = {1\over2} \lambda_{ij}(\qvec, \{ {\vec
  e_i} \}) u_i u_j$, where $ \{ {\vec e_i} \}$
characterize the orientation of the crystal lattice (e.g., they could
be a basis for the given lattice). By rotation average we mean the
average of $\lambda_{ij}(\qvec, \{ {\vec
 R e_i} \}) u_i u_j$ over all rotations $R$.
In practice, this amounts to using angular-averaged values for $S(p)$
and $n_p$ in Eq.~(\ref{eqfxcinftyij}).
Note that the above definition is equivalent to the average of
$\lambda_{ij}(R\qvec, \{ {\vec e_i} \}) (R{\vec u})_i (R{\vec u})_j$.
\end{thebibliography}
\end{document}